\documentclass[conference]{IEEEtran}
\usepackage[letterpaper, left=1in, right=1in, bottom=1in, top=0.75in]{geometry}

\usepackage{amsmath, amssymb, cite}
\usepackage{mathtools}
\usepackage[small]{caption}
\usepackage{amsthm,multirow,color,amsfonts}
\usepackage{tabulary}
\usepackage{subfigure}
\usepackage{graphicx}
\usepackage{setspace}
\usepackage{enumerate}
\usepackage{bbm}
\usepackage{comment}

\usepackage{setspace}	
\title{A Distributed Auction Policy for \\User Association in Device-to-Device\\ Caching Networks}

\author{
    \IEEEauthorblockN{Derya~Malak\IEEEauthorrefmark{1}, Mazin Al-Shalash\IEEEauthorrefmark{2} and Jeffrey~G.~Andrews\IEEEauthorrefmark{1}}
    \IEEEauthorblockA{\IEEEauthorrefmark{1}Department of Electrical and Computer Engineering \\The University of Texas at Austin, Austin, TX 78701, USA}
    \IEEEauthorblockA{\IEEEauthorrefmark{2}Huawei Technologies, Plano, TX 75075, USA
    \\ Email: deryamalak@utexas.edu, mshalash@huawei.com, jandrews@ece.utexas.edu}
    \thanks{Manuscript last revised: {\today}.}}

\newtheorem{theo}{Theorem}
\newtheorem{defi}{Definition}

\newcommand{\SINR}{\textrm{SINR}}

\DeclareMathOperator*{\sinr}{\sf SINR}

\DeclareMathOperator*{\mhc}{\sf MHC}
\DeclareMathOperator*{\Rdd}{R_{\sf D2D}}
\DeclareMathOperator*{\Rdds}{R_{\sf D2D}^2}
\DeclareMathOperator*{\tx}{\lambda_t}
\DeclareMathOperator*{\rx}{\lambda_r}
\DeclareMathOperator*{\Bid}{\mathcal{B}_{\phi}}

\DeclareMathOperator*{\BidS}{\mathcal{B}_{\phi,S}}

\DeclareMathOperator*{\dd}{\sf D2D}

\DeclareMathOperator*{\ppp}{\sf PPP}

\DeclareMathOperator*{\csma}{\sf CSMA}

\IEEEoverridecommandlockouts \IEEEpubid{\makebox[\columnwidth]{ 978-1-5386-3531-5/17/\$31.00~\copyright~2017 IEEE \hfill} \hspace{\columnsep}\makebox[\columnwidth]{ }}
\begin{document}

\maketitle
\begin{abstract}
We propose a distributed bidding-aided Mat\'{e}rn carrier sense multiple access ($\csma$) policy for device-to-device ($\dd$) content distribution. The network is composed of $\dd$ receivers and potential $\dd$ transmitters, i.e., transmitters are turned on or off by the scheduling algorithm. 
Each $\dd$ receiver determines the value of its request, by bidding on the set of potential transmitters in its communication range. Given a medium access probability, a fraction of the potential transmitters are jointly scheduled, i.e., turned on, determined jointly by the auction policy and the power control scheme. The bidding-aided scheduling algorithm exploits (i) the local demand distribution, (ii) spatial distribution of $\dd$ node locations, and (iii) the cache configurations of the potential transmitters. We contrast the performance of the bidding-aided $\csma$ policy with other well-known $\csma$ schemes that do not take into account (i)-(iii), demonstrate that our algorithm achieves a higher spectral efficiency in terms of the number of bits transmitted per unit time per unit bandwidth per user. 
The gain becomes even more visible under randomized configurations and requests rather than more skewed placement configurations and deterministic demand distributions.
\end{abstract}


\maketitle

\section{Introduction}
\label{intro}
Content caching is the key enabling design technique for offloading from the cellular infrastructure to decentralized device-to-device ($\dd$) communication. Caching aims to maximize the probability that the desired content can be found in a nearby device, i.e., the local hit rate. Due to potentially high density of devices, 
novel ways of scheduling concurrent $\dd$ transmissions are required in order to avoid interference and optimize the caching performance.


Power control is an effective approach to handle interference. Different power control algorithms to either optimize resource utilization for $\dd$ have been proposed in \cite{YuDopRibTir2011}, 
or to maximize the coverage probability of the cellular link as detailed in \cite{LeeLinAndHea2015}. Interference analysis in carrier sense multiple access ($\csma$) wireless networks is implemented in \cite{BusCheGor2009}. 
A synchronous peer-to-peer signaling and a concomitant scheduling protocol is designed in \cite{Wu2013FlashlinQ} that enables efficient channel aware spatial resource allocation and achieves significant gains over a $\csma$ system. 

Distributed solutions have been proposed for scalability and to improve different utility metrics. For example, a Gibbs sampling approach for scheduling to minimize the total interference and the delay is proposed in \cite{Kauffmann2007}, and to learn how to optimize the placement to maximize the cache hit rate of cellular networks is analyzed in \cite{ChatBlas2016}. 
Femtocaching using small cell access points, i.e., helpers, to minimize total delay is studied in \cite{Shanmugam2013}. 

Content placement and delivery should be jointly designed to maximize the offloading gain of $\dd$ caching \cite{CheYanXio2017}. 
Fair traffic association is required to balance the total load among the nodes. When the traffic demand and the location of caches are regular enough, the strategy of selecting the nearest cache can actually be close to optimal, as demonstrated in \cite{KroGioRen2016}. If the locations are not regular, load balancing can result in the maximum load of order $\Theta(\log \log n)$, where $n$ is the number of servers and requests, as shown in \cite{PouSiaSha2016}. This is an exponential improvement in a maximum load compared to the scheme which assigns each request to the nearest available replica. Our distributed solution is motivated from load balancing in the context of caching, which also captures the local demand popularity and cache configurations, unlike prior work.

We consider a spatial caching network in which the $\dd$ receivers and the potential transmitters are uniformly distributed. We assume the 
placement configuration of the potential transmitters as given. For this system model, we propose a totally distributed scheduling policy for the potential transmitter process by capturing the local demand profile of the receivers, the spatial distribution and the availabilities of the transmitters, with the objective of maximizing the spectral efficiency. 

Our model is an auction-based dynamic scheduling policy in which each receiver bids on the set of potential transmitters in its communication range. A fraction of the transmitters are jointly scheduled based on an on-off power control strategy given a medium access probability (MAP). The scheduling is not done uniformly at random, rather it depends on the cache configurations. The proposed solution captures (i) the  cache configurations, (ii) the signal-to-interference-and-noise-ratio ($\sinr$) coverage probability conditioned on the potential transmitter process, and (iii) the file popularity via the distribution of the local requests. We demonstrate the performance of our model for a given configuration in terms of the average rate per user under independent reference model (IRM) traffic, then test its robustness under different popularity profiles.

\section{System Model}
\label{model}
We envision a D2D caching network model in which the locations of the receiver process $\Phi_r$ and the potential transmitter process $\Phi$ are assumed to form realization of two independent homogeneous two-dimensional spatial Poisson point process (PPPs) with densities $\lambda_r$ and $\lambda_t$, respectively. 

We assume that the catalog size of the network is $M$ and $\mathcal{M}=\{1,\hdots, M\}$ denotes the set of files. Each transmitter has a cache of finite size $N < M$. Each receiver makes a file request based on a general popularity distribution over the set of the files. The document requests are modeled according to the Independent Reference Model (IRM), and the popularity distribution is modeled by the pmf $p_r(n)$, $n\in\mathcal{M}$.

We have the following additional assumptions. 
\begin{itemize}
\item Consider a snapshot of the set of $\dd$ nodes at a 
tagged time slot where a subset of the potential transmitters $\Phi$ simultaneously access the channel given a MAP $p_A$.
\item At a given snapshot, the cache configuration, i.e., the set of cached files, does not change.
\item Each receiver makes a request for one file randomly sampled from $p_r$, and can associate with any transmitter within its communication range.
\item A transmission is successful only if the received $\sinr$ is above the threshold $T$, given that the 
associated transmitter caches the desired file.
\end{itemize}

{\bf A high level summary.} The problem at a high level can be described as follows. Each receiver is allowed to communicate with any potential transmitter in its communication range and needs to choose a link. Receiver $u$ is associated with potential transmitter $x$, estimates the link $\sinr$ and bids on $x$ if the desired content is available in $x$'s cache. The values of the receiver bids are reported to potential transmitter $x$, and $x$ computes the cumulated sum of these variables taken on all users in its cell. The potential transmitter $x$ then reports the value of the bid sum to other potential transmitters in its contention range. Given the accumulated bids of all potential transmitters, the exclusion (or contention) range\footnote{If a transmitter has other transmitters in its contention domain, its channel capacity will be a fraction of the medium capacity due to sharing of resources.} and the MAP, the algorithm determines the set of active transmitters.

Let $\tilde{\Phi}=\{(x,m_x,{\bf P}_x)\}$ be an independently marked PPP with intensity $\lambda_t$, where i) $\Phi=\{x\}$ denotes the locations of potential transmitters, ii) $\{m_x\}$ are the marks of $\tilde{\Phi}$, and iii) ${\bf P}_x=(P_x^y: y)$ denotes the virtual power\footnote{Virtual power $P_x^y$ is the product of the effective power of transmitter $x$ and of the random fading from this node to receiver $y$.} emitted by node $x$ to node $y$ provided it is authorized by the MAC mechanism. The random variables ${\bf P}_x$ are iid, exponential with mean $\mu^{-1}$.

\begin{defi}{\bf Neighborhoods.}
The neighborhood system on $\Phi$ is the family $N=\{\mathcal{N}(x)\}_{x\in \Phi}$ of subsets of $\Phi$ such that for all $x\in \Phi$, we have $x\notin \mathcal{N}(x)$, and $z\in \mathcal{N}(x) \implies x\in \mathcal{N}(z)$. The subset $\mathcal{N}(x)$ is called the neighborhood of node $x$. 
\end{defi}

\begin{table}[t!]\footnotesize
\begin{center}
\setlength{\extrarowheight}{3pt}
\begin{tabular}{l | c }
{\bf Symbol} & {\bf Definition} \\ 
\hline
$\ppp$ dist. $\dd$ receivers; potential transmitters & $\Phi_r$; $\Phi$ \\
Medium access prob.; set of active transmitters & $p_A$; $\Phi_t$\\
Density of $\Phi_r$; density of $\Phi$ & $\rx$; $\tx$\\
$\sinr$ threshold; noise power at the receiver & $T$; $\sigma^2$\\ 
Ball centered at node $x$ with radius $R$ & $B_x(R)$\\
$\dd$ radius; exclusion radius for Mat\'{e}rn $\csma$ & $\Rdd$;  $D$\\
File request distribution & $p_r\sim \rm{Zipf}(\gamma_r)$\\
Total number of files; cache size; set of all files & $M; N$; $\mathcal{M}$\\ 
File requested by $u\in \Phi_r$; cache config. of $x\in \Phi$ & $c_u$; $\mathcal{C}_x$\\
Power law path loss function & $l(r)=r^{-\alpha}$\\
Accumulated bid of transmitter $x$ & $\Bid(x)$\\
\end{tabular}
\end{center}
\caption{Notation.}
\label{table:tab1}
\end{table}

For $x\in\Phi$, let the neighbors of node $x$ be
\begin{eqnarray}
\mathcal{N}(x)=\{(y,m_y,{\bf P}_y)\in\tilde{\Phi}: P_y^x/l(|x-y|)\geq P_0, y\neq x\},\nonumber
\end{eqnarray}
i.e., the nodes in its contention domain. If we only consider path-loss and no fading, the received signal at the boundary should be larger than the threshold, equivalent to $D=(\mu P_0)^{-1/\alpha}$ for a fixed transmit power of $\mu^{-1}$. Thus, $P_y^x/l(|x-y|)\geq P_0$ will be equivalent to $y\in B_x(D)$, where $B_x(D)$ is a ball centered at $x$ with contention radius $D$.


The medium access indicators $\{e_x\}_x$ are additional dependent marks of the points of $\Phi$ as follows:
\begin{align}
\label{exdefinition}
e_x=\mathbbm{1}\left(\forall_{y\in \mathcal{N}(x)} m_x<m_y\right).
\end{align}

The set of transmitters retained by $\csma$ as a non-independent thinning of the PPP $\Phi$, and denoted by 
\begin{align}
\Phi_t= \{x\in\Phi\vert e_x=1\}.
\end{align}
The probability of medium access of a typical node equals $p_A=\mathbb{E}^0[e_x]$, where $\mathbb{E}^0$ is the expectation with respect to $\Phi$'s Palm probability $\mathbb{P}^0$; i.e., $\mathbb{P}^0(\Phi(\{0\})\geq1) = 1$ \cite[Ch. 4]{BaccelliBook1}.

Next, by incorporating the $\sinr$ coverage characteristics in a realistic $\dd$ network setting with contention prevention provided by the $\mhc$-II model, we envisage a bidding-aided scheduling policy in Sect. \ref{BiddingApproach}.

\section{Bidding-Aided Policy for 
Associations}
\label{BiddingApproach}
Using the potential transmitter model just described, the potential received $\sinr$ of a receiver located at $z$ covered by $x\in\Phi$ is expressed as
\begin{align}
\label{SINRrigorous}
\SINR_x(u)=\frac{P_{xu}l(|x-u|)}{\sigma^2+\sum\nolimits_{z\in \Phi\backslash \{x\}}P_{zu}l(|z-u|)},
\end{align}
where $r=|x-u|$ is the distance between the potential transmitter located at $x\in\Phi$ and the receiver $u$, and for a fixed path-loss exponent $\alpha$, $l(r)=r^{-\alpha}$ under OPL3 \cite[Ch. 2.3]{BaccelliBook1}, and $r$ and $r_z=|z-u|, z\in \Phi$ denote the distance between the potential transmitter and the receiver, and the interferers and the receiver, respectively, and $\sigma^2$ is the noise power at the receiver side. Similarly, $\{P_{zu}\}_{z\in \Phi}$ 
are random variables that denote the on-off powers of potential transmitters, i.e., 
\begin{align}
P_{zu}=1_{z\in\Phi_t},
\end{align}
where $\Phi_t$ is a repulsive 
point process that models the retained process of transmitters. The procedure to decide the set of retained 
transmitters will be detailed next.

We develop a bidding-based user association algorithm such that receivers are associated in a way to maximize the ``local cache hit probability". We introduce an on-off distributed power control method with coordination between the neighboring transmitters for the $\dd$ caching framework\footnote{On-off power control requires the CSI knowledge about the direct link between the transmitter and its corresponding receiver \cite{LeeLinAndHea2015}. We 
only consider long term CSI (ignore fading).}. For a fixed probability of medium access\footnote{Only a certain fraction of transmitters is to be activated 
to control interference and provide the $\dd$ users with high spectral efficiency.}, the bidding algorithm 
determines which links to activate by capturing the matchings between the availability of the caches and the local demand. 

Each receiver $u\in \Phi_r$ bids on the potential transmitters $x\in\Phi$ in its range $\Rdd$ based on their virtual $\sinr$ coverage probability characteristics. Each $x\in\Phi$ accumulates bids from the receivers conditional on the cache configuration. Because the local demand and the coverage characteristics will be similar, the transmitters located at similar geographic locations collect similar bids. Upon the assignment of the bids of all the potential transmitters, $x$ is scheduled if it has the highest bid inside a circular exclusion region $B_x(D)$. 
Hence, the process of retained transmitters $\Phi_t$ will be obtained as a dependent thinning of $\Phi$, in contrast with the Mat\'{e}rn hard-core (MHC) model where the potential transmitters are assigned iid marks. We next discuss the technical details of the bidding approach.

\subsection{Accumulated Bid of a Potential Transmitter}
For given realizations $\phi$ of $\Phi$, and $\phi_r$ of $\Phi_r$, the total bid collected at a potential transmitter $x\in \phi$ is determined using the following expression:
\begin{align}
\label{bidx}
\Bid(x)
=\sum\limits_{u\in \mathcal{U}_x}{p_r^x(c_u)\mathbb{P}({\SINR}_x(u)>T}),\, x\in\phi,
\end{align}
where for the general coverage model with noise and interference, we denote by 
\begin{align}
\mathcal{U}_x=\{u\in \phi_r\cap B_x(\Rdd)\vert x\in \phi, c_u\in\mathcal{C}_x\}
\end{align}
is the set of receivers bidding on potential transmitter $x$. 

Note that (\ref{bidx}) is  a weighted sum of the virtual $\sinr$ coverage distributions of the set of receivers inside the coverage region with radius $\Rdd$ of the potential transmitter $x$. The parameter $c_u$ (sampled iid from $p_r$) denotes the index of the file requested by receiver $u$, and $\mathcal{C}_x$ denotes the set of files available in the cache of transmitter $x\in \Phi$, i.e., the cache configuration of $x$. The local request distribution observed at $x\in\phi$, i.e., the request distribution conditioned on the cache configuration of $x\in\phi$, 
is given as
\begin{align}
\label{localrequestsconditioned}
p_r^x(m)={|\mathcal{U}_x(m)|}/{{|\mathcal{U}_x|}},\quad x\in\phi, \,\, m\in \mathcal{C}_x,
\end{align}
where 
\begin{align}
|\mathcal{U}_x(m)|&=\sum\nolimits_{u\in \phi_r\cap B_x(\Rdd)}  {\mathbbm{1}(c_u=m)\mathbbm{1}(m\in\mathcal{C}_x)},\quad x\in \phi \nonumber\\
|\mathcal{U}_x|&=\sum\nolimits_{u\in \phi_r\cap B_x(\Rdd)\vert x\in \phi}  \mathbbm{1}(c_u\in\mathcal{C}_x)
\end{align}
are the number of receivers in the coverage of $x$ that request file $m\in\mathcal{C}_x$, and the cardinality of the set of users associated to $x\in \phi$, respectively. 

\begin{figure*}[t!]
\centering
\includegraphics[width=0.45\textwidth]{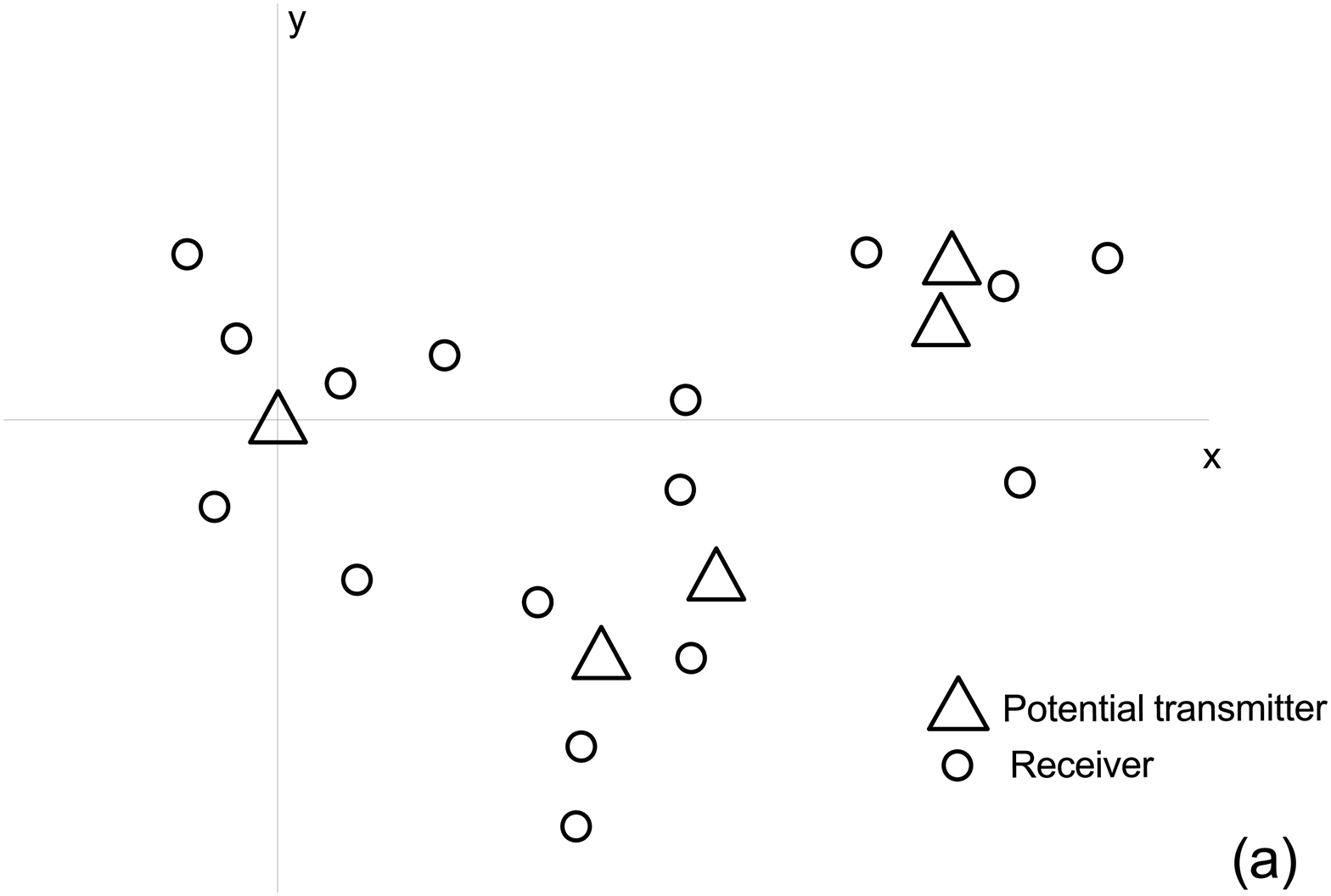}
\includegraphics[width=0.45\textwidth]{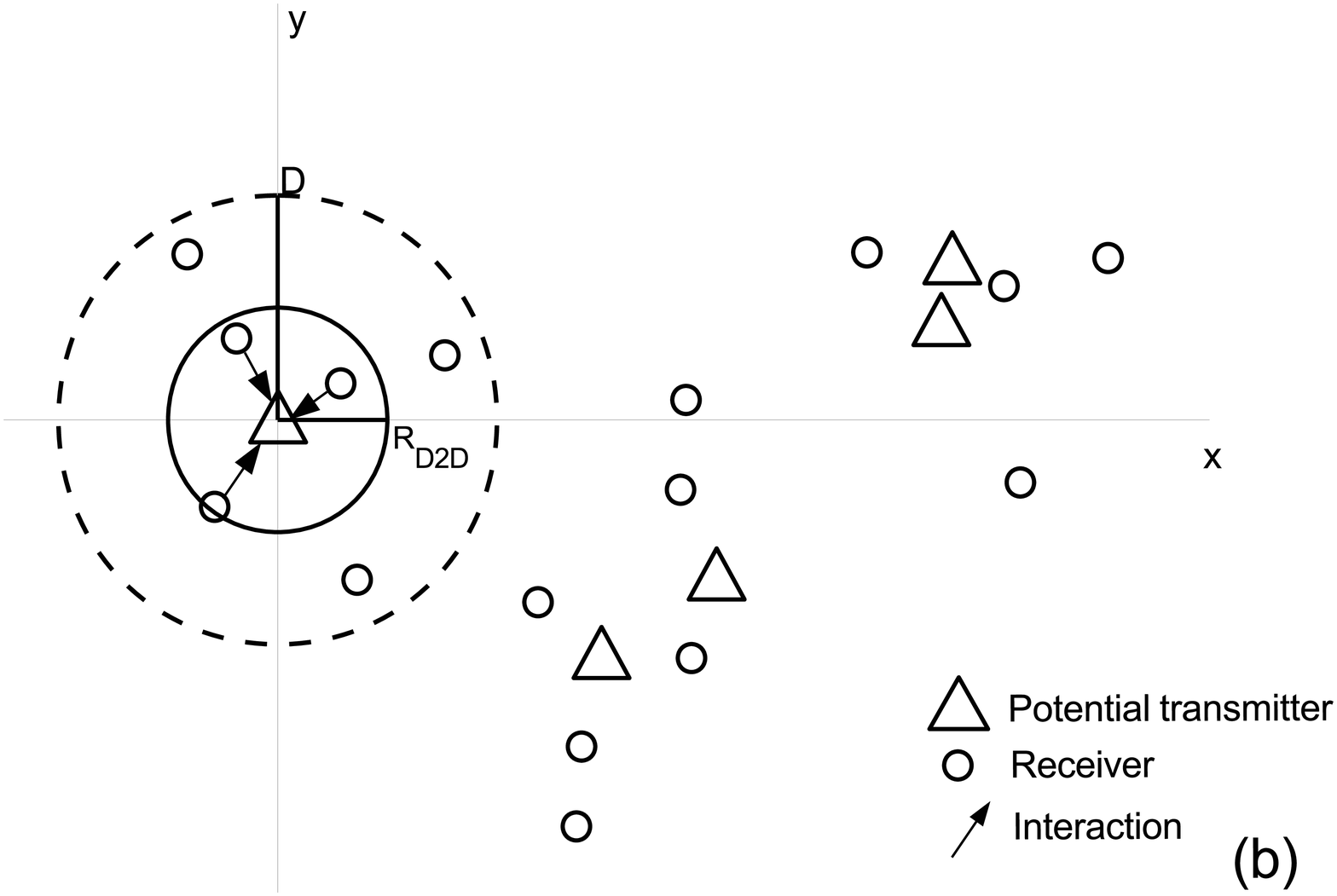}
\includegraphics[width=0.45\textwidth]{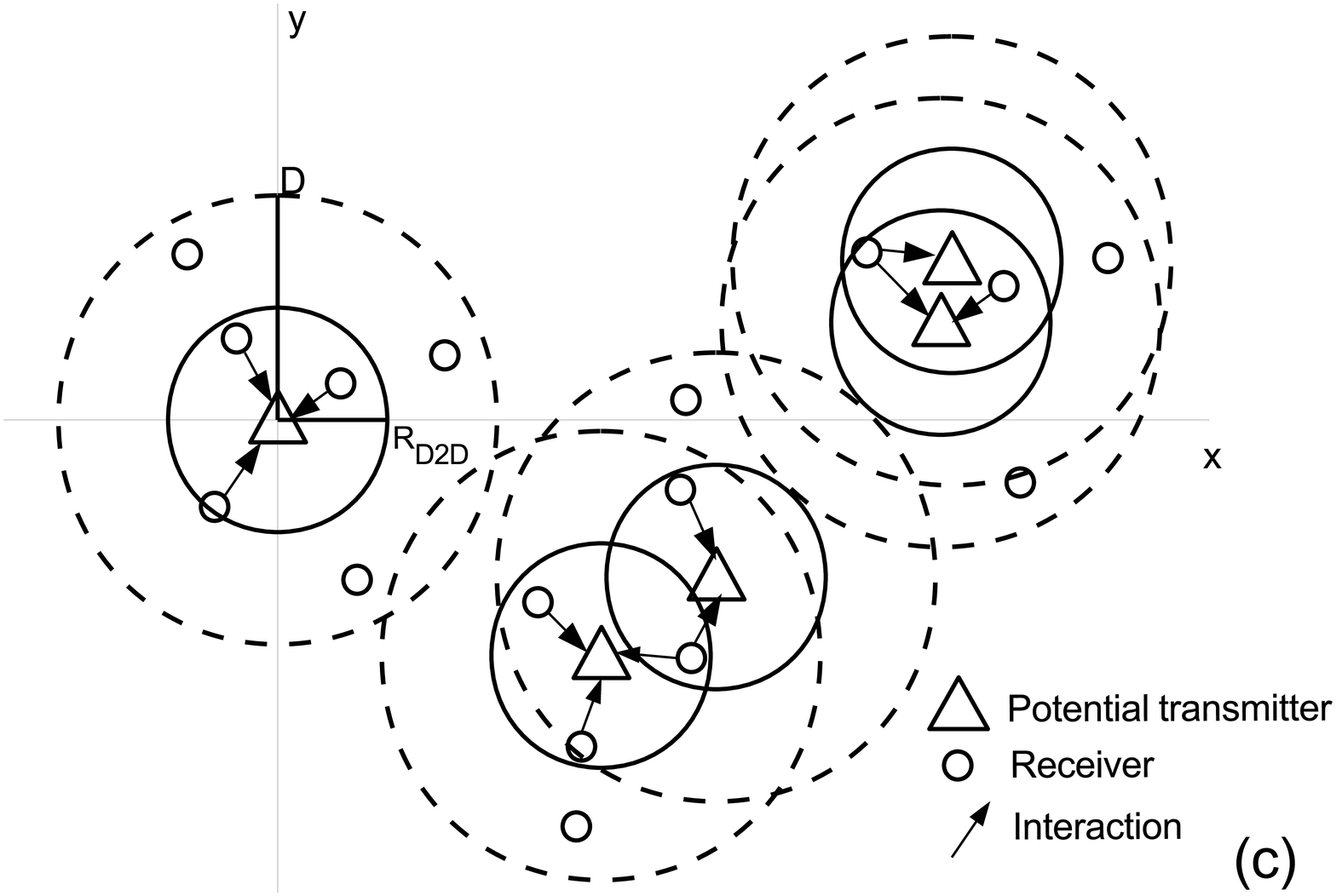}
\includegraphics[width=0.45\textwidth]{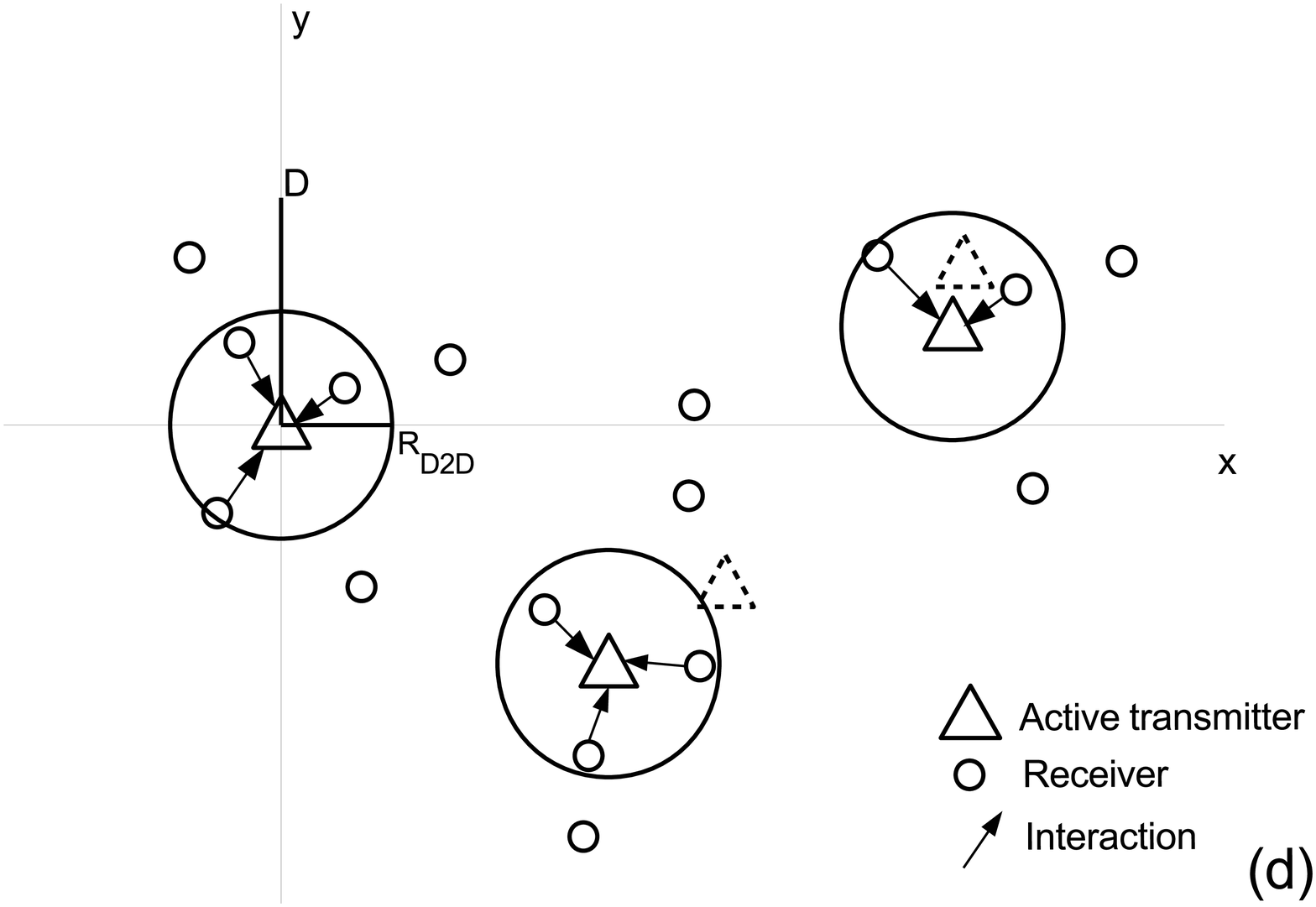}
\caption{\small{A visualization of the bidding algorithm on the receiver and the potential transmitter processes. Consider the network setup in Fig. \ref{BiddingAlgo}-$(a)$ with the set of potential transmitters and receivers. Fig. \ref{BiddingAlgo}-$(b)$ shows the interactions between the potential transmitter centered at origin, where the solid (dashed) circle shows the communication (exclusion) range. A receiver can bid on the potential transmitter only if it is in the communication range. Fig. \ref{BiddingAlgo}-$(c)$ shows the system-level interactions that might overlap depending on the potential transmitter locations. Fig.\ref{BiddingAlgo}-$(d)$ shows the set of retained transmitters.}
\label{BiddingAlgo}}
\end{figure*}

The bidding formulation in (\ref{bidx}) captures the 
\begin{itemize}
\item cache availability via 
conditioning on the set $\mathcal{U}_x$,
\item $\sinr$ coverage conditioned on the potential transmitter process $\phi$, and 
\item file popularity through the local request distribution $p_r^x$ as defined in (\ref{localrequestsconditioned}).
\end{itemize}
Using this bidding formulation, we 
analyze the bidding algorithm in Sect. \ref{HomogeneousPPPbidding} to determine the set of retained transmitters $\phi_t$. We illustrate the 
algorithm in Fig. \ref{BiddingAlgo}.

Similar to a hard-core process, $\phi_t$ has an exclusion radius of $D$ possibly different from the communication radius $\Rdd$ that will be determined in Sect. \ref{Rd2dvsD}. 

The cardinality of receivers that bid on $x\in\phi$, i.e., $|\mathcal{U}_x|$, is distributed as ${\rm Poisson}(\lambda_r^x\pi \Rdds)$, where the intensity of receivers that bid on transmitter $x$ is given by $\lambda_r^x=\lambda_r\sum\nolimits_{m\in\mathcal{C}_x}{p_r(m)}$. Hence, the average number of receivers associated to $x\in \phi_t$ is 
$\mathbb{E}[|\mathcal{U}_x|]=\lambda_r^x\pi \Rdds$, and the distribution of $|\mathcal{U}_x|$ satisfies
\begin{align}
\mathbb{P}(|\mathcal{U}_x|=n)&=\exp{(-\lambda_r^x\pi \Rdds)}\frac{(\lambda_r^x\pi \Rdds)^n}{n!}.
\end{align}

The rest of this section is mainly devoted to the special case of the homogenous PPP approximation for the bidding algorithm and its distributional characteristics. Note that the PPP assumption here does not imply that the thinning of $\Phi$ is done independently. Instead, as will be shown next, it yields a dependent thinning of $\Phi$.

\subsection{Analysis of Bidding Model}
\label{HomogeneousPPPbidding}
The process of transmitters $\Phi_t$ arranged according to some homogeneous PPP of intensity $\lambda=p_A\tx$ in the Euclidean plane. For the general $\sinr$ regime, the probability of coverage of a typical randomly located receiver in the general cellular network model, where the transmitters are arranged according to some homogeneous PPP is evaluated in \cite{Andrews2011}. The coverage probability of a user $u$ (assuming 
nearest transmitter association)
is 
\begin{align}
\label{SINRdistributiongivendistance}
\mathbb{P}[{\SINR}>T]=e^{-\mu T\sigma^2 /l(r)}\mathcal{L}_{I_r}(\mu T /l(r)),
\end{align}
where $r$ denotes the distance from the receiver to the serving transmitter, and $\mathcal{L}_{I_r}(s)$ is the Laplace transform of the interference and is given by
\begin{align}
\mathcal{L}_{I_r}(s)=\exp\Big(-\pi \lambda\int\nolimits_{r^2}^{\infty}{\frac{1}{1+\mu s^{-1}t^{\alpha/2}}{\rm d}t}\Big).\nonumber
\end{align}
Hence, we can compute $\mathcal{L}_{I_r}(\mu T /l(r))$ as
\begin{align}
\mathcal{L}_{I_r}(\mu T /l(r))
\overset{(a)}{=}\exp\big(-\pi \lambda \rho(T,\alpha) r^2\big),
\end{align}
where $(a)$ follows from employing a change of variables $z=t/(T^{2/\alpha} r^2)$, where $\rho(T,\alpha)=T^{2/\alpha}\int\nolimits_{T^{-2/\alpha}}^{\infty}{\frac{1}{1+z^{\alpha/2}}{\rm d}z}$, 

Cumulated bid (\ref{bidx}) of potential transmitter $x\in\phi$ can be rewritten using the $\sinr$ distribution given in (\ref{SINRdistributiongivendistance}) as
\begin{multline}
\label{PPP_SINR_bid}
\Bid(x)=\sum\limits_{u\in \mathcal{U}_x}{p_r^x(c_u)e^{-\mu T\sigma^2 /l(r_{xu})}\mathcal{L}_{I_{r_{xu}}}(\mu T /l(r_{xu}))}\\
=\sum\limits_{u\in \mathcal{U}_x}{p_r^x(c_u)\exp{\left(-\mu T\sigma^2 /l(r_{xu})-\pi\lambda\rho(T,\alpha)r_{xu}^2\right)}}\\
\overset{(a)}{=}\sum\limits_{u\in \mathcal{U}_x}{p_r^x(c_u)\left(1-\mu T\sigma^2 /l(r_{xu})-\pi\lambda\rho(T,\alpha)r_{xu}^2\right)},
\end{multline}
where $r_{xu}=|x-u|$, and 
$(a)$ is required for analytical tractability. 
The total bid expression in (\ref{PPP_SINR_bid}) is a random variable as a function of the local request distribution $p_r^x(c_u)$ of $u\in \mathcal{U}_x$. Conditioning on the value of $|\mathcal{U}_x|$, $u\in \mathcal{U}_x$ are iid and uniformly 
distributed within 
$B_x(\Rdd)$.

The spatial distribution of the bids can be calculated using a similar approach to the one proposed in \cite{Ganti2009}. 

\subsection{Communication Range versus Exclusion Range}
\label{Rd2dvsD}
Given a contention-based model, the interference measured at the typical point depends on the range of the contention domain. Hence, the range at which the communication is successful, i.e., $\SINR\geq T$, is determined by the exclusion radius. 
Using the $\sinr$ expression in (\ref{SINRrigorous}), we rewrite the $\sinr$ for noise- and interference-limited regimes as follows:
\begin{align}
\label{SINRlimited}
\SINR=\begin{cases}
{h l(r)}/{\sigma^2},\,\, I\to 0,\\
{h l(r)}/{\bar{I}},\,\, \sigma^2\to 0,
\end{cases}
\end{align}
respectively, where $h$ is the exponential channel gain with parameter $\mu$. The communication range is defined by $\Rdd$ such that $r\leq \Rdd \implies \SINR\geq T$.

Using (\ref{SINRlimited}), and neglecting the small scale Rayleigh fading variability, it is easy to note that in the noise-limited regime, there is a one-to-one mapping between $T$ and $\Rdd$. Unlike the noise-limited regime, $\Rdd$ for the interference-limited regime is variable. To ease the analysis in the interference-limited regime, we approximate the interference $I$ by its mean $\bar{I}$. Hence, one can derive the communication range
\begin{align}
\Rdd =
\begin{cases}
(\mu T\sigma^2)^{-1/\alpha},\,\, I\to 0,\nonumber\\
(\mu T\bar{I})^{-1/\alpha},\,\, \sigma^2\to 0.\nonumber
\end{cases}
\end{align}

We benefit from a very useful approximation to characterize $\bar{I}$, which is first suggested in \cite{Haenggi2011}. The excess interference ratio (EIR) as defined in \cite{Haenggi2011} is the mean interference measured at the typical point of a stationary hard-core point process of intensity $\lambda$ with minimum distance $D$ relative to the mean interference in a Poisson process of intensity $\lambda(r) =\lambda{\bf 1}_{[D,\infty)}(r)$. Their analysis shows that the excess interference ratio for Mat\'{e}rn processes of type II ($\mhc$-II) never exceeds $1$ dB. Thus, using a modified path loss law $\tilde{l}(r)=l(r){\bf 1}_{r>D}$, the mean interference is approximated as 
\begin{align}
\bar{I}\approx\lambda \int\nolimits_{\mathbb{R}^2}{\tilde{l}(|y|){\rm d}y}
=2\pi\lambda \int\nolimits_{D}^{\infty}{r^{-\alpha+1}\,{\rm d}r}=\frac{2\pi\lambda}{\alpha-2}D^{2-\alpha}, \nonumber
\end{align}
using which $\Rdd$ can be approximated as a function of the exclusion radius $D$ as the interference varies.

\section{Process of Retained Transmitters}
\label{RetainedProcess}
Let $\{m_x\}$ be random variables (marks) over $x\in\Phi$ that are iid and uniformly distributed on $[0,1]$. 
Consider the following scheduling policies:

\paragraph{Random selection} In this model, each transmitter is randomly activated with probability $p_A$, where there is no exclusion region around the transmitters. This case is equivalent to assigning marks $\{m_x\}$ to $x\in\tilde{\Phi}$. Thus, the medium access indicator of node $x$ is 
\begin{align}
e_x^R=\mathbbm{1}\left(m_x<p_A\right).
\end{align}

\paragraph{Mat\'{e}rn $\csma$}  In the case of $\mhc$ thinning, the potential transmitters $x\in\tilde{\Phi}$ are assigned marks $\{m_x\}$, and a transmitter is retained if it has the ``lowest mark" or ``highest mark" within the exclusion region. Hence, we have
\begin{align}
e_x^M=\mathbbm{1}\left(\forall_{y\in \mathcal{N}(x)} m_x<m_y\right),
\end{align}
where the parameter $D$ is determined using the first-order characteristics $p_A\lambda_t$. 

\paragraph{Bidding-aided Mat\'{e}rn $\csma$}
Consider the following bidding-aided Mat\'{e}rn $\csma$ thinning model, where instead of assigning iid and uniformly distributed marks $\{m_x\}$ on $[0,1]$ to each of $x\in\tilde{\Phi}$, we compare the cumulated bid values $\{\Bid(x)\}_x$ and retain the transmitters that have the highest bid value within the exclusion region. Hence, we have
\begin{align}
e_x^B=\mathbbm{1}\left(\forall_{y\in \mathcal{N}(x)}  \Bid(x)>\Bid(y)\right).
\end{align}

\paragraph{Bid ordering} In this scheme, given a realization $\phi$ of $\Phi$ with cardinality $|\phi|=N$, bids are sorted in descending order. The sorted bid vector is given as $\BidS={\sf sort}_{x\in\Phi}(\Bid(x))$. For a given probability of medium access $p_A$, node $x$ is retained if its bid rank is at most $\lfloor p_AN \rfloor$. The medium access indicator is 
\begin{align}
e_x^O=\mathbbm{1}\left(\Bid(x)\geq \BidS(\lfloor p_A N \rfloor)\right).
\end{align}

\begin{figure*}[t!]
\centering 
\includegraphics[width=.495\textwidth]{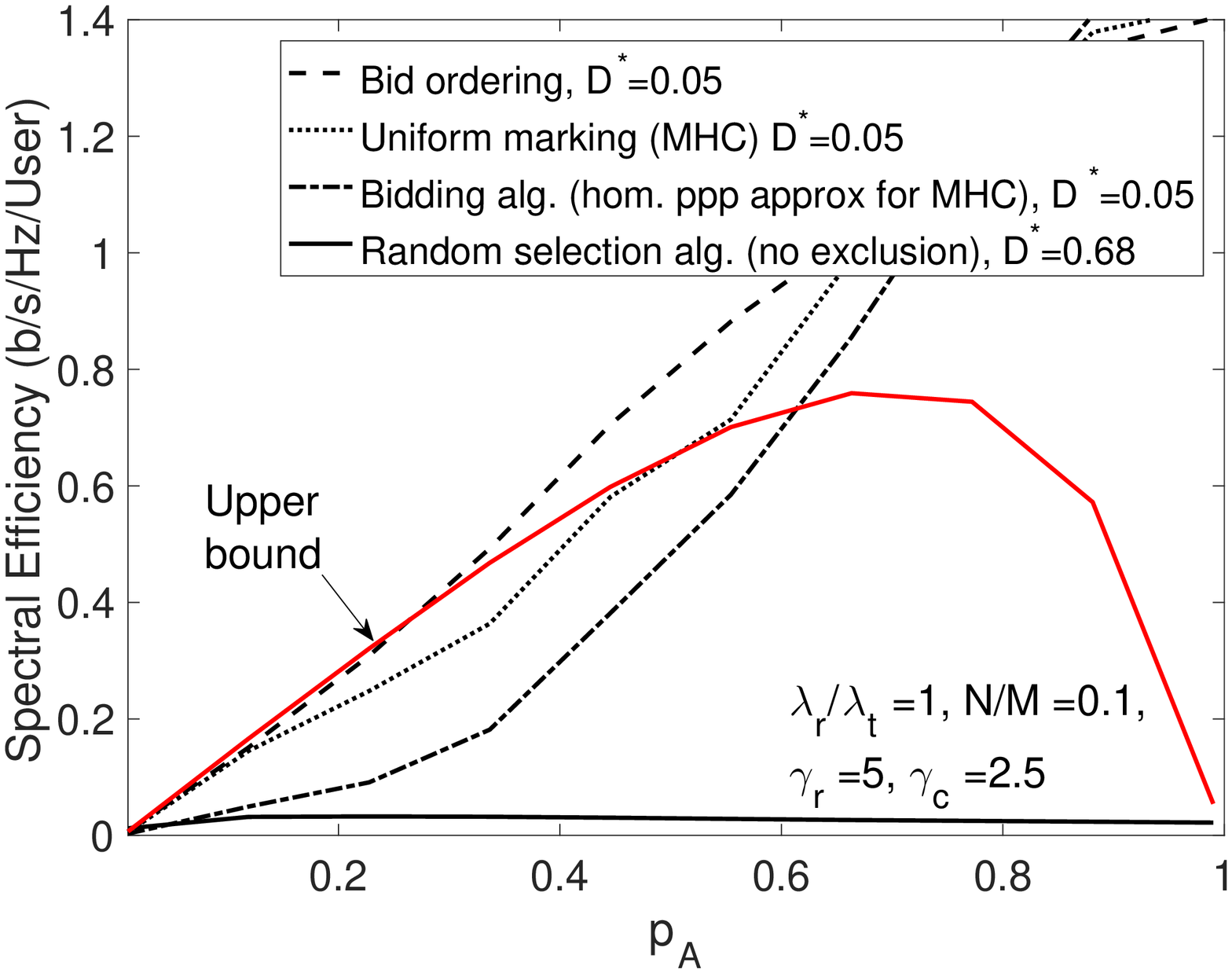}
\includegraphics[width=.495\textwidth]{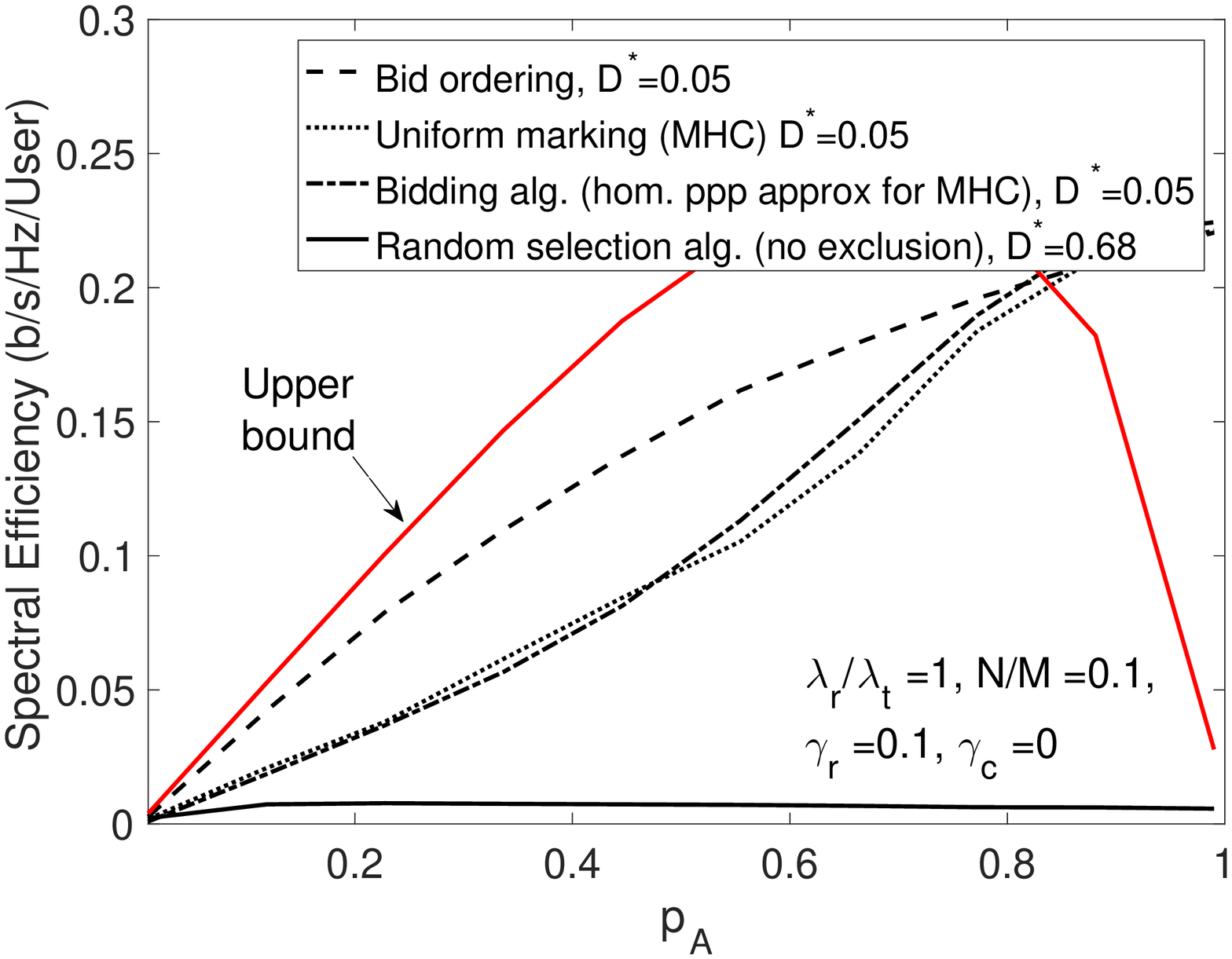}
\caption{\small{Spectral efficiency comparison of the bidding-aided $\csma$ model with other scheduling policies. (Left) Skewed cache configurations and requests. (Right) Randomized cache configurations and requests.}\label{spectralefficiency}}
\end{figure*}

{\bf Spectral Efficiency.} 
Spectral efficiency gives the number of bits transmitted per unit time per unit bandwidth. For tractability, we assume that each transmitter allocates equal time-frequency resources to its users, i.e., each user gets rate proportional to the spectral efficiency of its downlink channel from the serving transmitter. For total effective bandwidth $W$ Hz, the average downlink rate in bits/sec of a typical user is
\begin{align}
\label{SE}
\mathbb{E}[R\vert N>0]=\mathbb{E}\left[\frac{W}{\tilde{N}}\log_2(1+{\SINR})\mathbbm{1}_{\SINR\geq T}\right],
\end{align}
where $N$ is the number of users served by the tagged transmitter, and $\sinr$ is characterized by incorporating Rayleigh fading. The distribution of $N$ (for the $\ppp$ BS setting) is characterized in \cite{SinDhiAnd2013}. 
Given there is at least one user associated to the tagged transmitter, which occurs with probability $\mathbb{P}(N>0)=1-\exp(-\Lambda_r)$, where $\Lambda_r=\lambda_r\pi \Rdds$ is the average number of receivers in the communication range of the transmitter, the conditional probability of having $N=k$ receivers is given as
\begin{align}
\mathbb{P}(\tilde{N}=k)=\frac{\Lambda_r^k \exp(-\Lambda_r)}{k!(1 - \exp(-\Lambda_r ))}.\nonumber
\end{align}
Our objective in Sect. \ref{performance} is to compare the bidding-aided $\csma$ policy 
with the other popular algorithms 
summarized above using the spectral efficiency 
in (\ref{SE}).

\section{Performance Evaluation}
\label{performance}
We consider a realization $\phi$ of PPP $\Phi$ over the region $S=[-5,5]^2$ with an intensity of $\lambda_t=3$ 
per unit area. The catalog size is $M=100$ files and each potential transmitter $x\in\phi$ can store up to $N=10$ files. We consider an IRM traffic scenario, where the popularity of requests is modeled by the Zipf distribution, which has pmf $p_r(n) =\frac{1}{n^{\gamma_r}}\Big/\sum\nolimits_{m=1}^M{\frac{1}{m^{\gamma_r}}}$, for $n\in\mathcal{M}$, where $\gamma_r\in (0,1)$ the Zipf exponent that determines the skewness of the distribution. File requests are generated over $S$ according to a time and space homogeneous $\ppp$ with intensity $\lambda_r =3$ requests per unit time per unit area, and file requests are uniform and independent over the space, and any new request can be for $m \in \mathcal{M}$ with probability $p_r(m)$. The rest of the network parameters are chosen as follows. Path loss exponent is $\alpha = 4$, $\sinr$ threshold is $T = 0.01$, $\sigma^{-2} = .1$, and the fading parameter is $\mu = 1$. 

Next, we illustrate the performance of different scheduling algorithms as a function of the MAP $p_A$. In Fig. \ref{spectralefficiency}-$(a)$, we have a skewed placement configuration $p_c\sim$ Zipf($2.5$) and $p_r\sim$ Zipf($5$). In Fig. \ref{spectralefficiency}-$(b)$, we have $p_c\sim$ Zipf($0$) and $p_r\sim$ Zipf($0.1$). We also compare against analytical upper bounds using the modified MHC 
model proposed in \cite{YeShaCarAnd2013}, and an upper bound for the low contention regime of $\csma$ using \cite[Ch. 3.7]{HaenggiGanti2009}. The bidding algorithm provides higher throughputs than random selection and uniform marking. For skewed placement, the spectral efficiency performance is very close to the upper bound for the low contention regime.



\section{Conclusions}
\label{conc}
We developed a bidding-aided distributed scheduling policy for $\dd$ users by capturing the local demand profile, the spatial distribution and the configurations of the transmitters, with the objective of maximizing the spectral efficiency in the units of bits/sec/Hz/User. We demonstrated and contrasted the performance of our bidding-aided algorithm with other well-known $\csma$ policies. The key takeaways include that rather than solely balancing the traffic according to the locations of caches, exploiting the cache configurations and local demand distribution, higher throughput gains can be achieved, and our approach provides new insights into designing dynamic bidding-aided caching algorithms. Possible directions include the extension of the scheduling algorithm to develop dynamic caching algorithms that capture the network configuration in order to achieve higher throughput scaling gains with caching.

\begin{spacing}{0.81}
\bibliographystyle{IEEEtran}
\bibliography{D2Dreferences}
\end{spacing}

\end{document}